# Electrically and Magnetically Resonant dc SQUID Metamaterials.

O.V. Shramkova[1], N. Lazarides[1,2,3], G.P. Tsironis[1,2,3] and A.V. Ustinov[2,4]

[1]Crete Center for Quantum Complexity and Nanotechnology, University of Crete, Heraklion, Greece
[2]National University of Science and Technology, MISiS, Moscow, Russia
[3]Institute of Electronic Structure and Laser, Foundation for Research and Technology–Hellas, Heraklion, Greece.
[4]Physikalisches Institut, Karlsruhe Institute of Technology, Karlsruhe, Germany
*corresponding author, E-mail: oksana@physics.uoc.gr

## Abstract

We propose a superconducting metamaterial design consisting of meta-atoms (MAs) which are each composed of a direct current (dc) superconducting quantum interference device (SQUID) and a superconducting rod. This design provides negative refraction index behavior for a wide range of structure parameters.

## 1. Introduction

Artificial electromagnetic media such as metamaterials (MM) have been extensively studied during the last decade. Negative-index metamaterial (NIM) is a metamaterial that exhibits negative effective permittivity and permeability over some frequency range. Such a medium has a negative index of refraction [1]. MMs manufactured by the modern microfabrication and nanofabrication techniques are comprised of the periodic arrangements of unit cells, which are significantly smaller than the wavelength of electromagnetic wave. The properties of MMs are determined by both the individual constituent elements and collective response of their ensembles. The initial experimental verification of a negative refractive index was reported for NIMs composed of arrays of metallic split-ring resonators (SRRs) and continuous wires [2],[3]. However, high Ohmic losses of a metal can ruin negative refraction [4], [5]. To provide the loss reduction and wideband tunability in MMs we can replace the metallic elements by the superconducting ones [6],[7]. Temperature, current and magnetic field can be used for frequency tuning of superconducting MMs, which is attractive for different applications. The replacement of metallic SRRs with radio frequency (rf-) SQUIDs was suggested theoretically in [8]. The rf-SQUID consists of a superconducting ring interrupted by a Josephson junction (JJ). The JJ shows the nonlinear current-voltage characteristic due to microscopic phenomenon. The electromagnetic properties of SQUID-based MM are determined by the resonant characteristics of the individual constitutive elements and rely on the magnetic coupling of its elements through dipole-dipole forces due to the mutual inductance between SQUIDs. The tunability and reconfigurable nature of rf-SQUID MMs was investigated theoretically and experimentally [9]-[17].

The direct current (dc) SQUID is formed by a ring with two JJs. A sketch of such a device is presented in Fig.1 (a). An equivalent circuit that can be used for this type of SQUID is shown in Fig.1 (b) and will be described below. If the total flux through the loop is $n\Phi_0$ ($n=0,1,2...$), two links have the same phase difference, which results in constructive interference. For the flux corresponding to $\left(n+\frac{1}{2}\right)\Phi_0$ the links introduce opposite phase difference leading to destructive interference. The dc-SQUID is an extremely sensitive detector of magnetic flux. This sensitivity of the SQUID is utilised in many different fields of applications, including biomagnetism, materials science, metrology, astronomy and geophysics [18],[19].

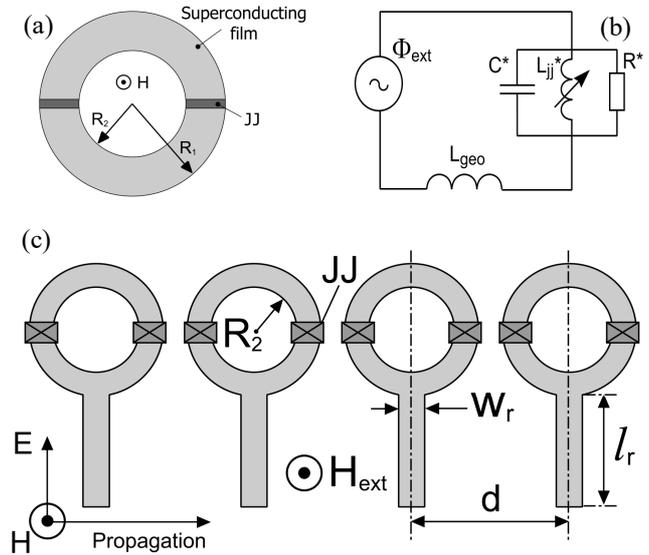

Figure 1: (a) – Schematic drawing of a dc-SQUID, (b) – Equivalent electrical circuit for a single SQUID, (c) – 1D SQUID metamaterial.

In the present work we investigate the response of one-dimensional (1D) MMs that employs dc-SQUIDs with attached superconducting rods as MAs. In the Section 2 we present a brief overview of the main equations. The mathematical formulation is based on the circuit approach. In the Section 3 we give the results of numerical simulations. The effect of the various geometrical parameters on the left-handed (LH) behavior of the structure is also considered.

## 2. Modeling

Let us consider the MM with MAs comprised of dc-SQUIDs with superconducting rods (Fig.1(c)). The time-dependent magnetic field **H**(t) is applied perpendicular to the plane of the SQUID loop. Junctions are symbolized as the dark crosses. The electromagnetic field, whose magnetic component is perpendicular to the SQUID loop and whose electric component is directed along the rod (TEM mode), is applied.

### 2.1. Effect of dc-SQUIDs

The geometric inductance of the two arms is small compared to the Josephson inductance. As the result, the dc-SQUID can be considered like a single junction with a tunable effective critical current. The equivalent lumped circuit for the dc-SQUID in a magnetic field comprises a flux source $\Phi_{ext}$ in series with the geometric inductance of the SQUID loop and JJ shunted by effective capacitor $C^*$ ( $C^* = C_1 + C_2$ ) and effective resistor $R^*$ ( $\frac{1}{R^*} = \frac{1}{R_1} + \frac{1}{R_2}$ ), where JJs are shunted by resistance $R_{1,2}$ and capacitance $C_{1,2}$ (Fig.1(d)). Here and below subscripts 1 and 2 correspond to the different JJs.

If the two links have the phase difference $\delta_1$ and $\delta_2$ respectively, the total current producing as a result of Josephson effect will be

$$I_{jj} = I_0 \sin\delta_1 + I_0 \sin\delta_2, \quad (1)$$

where $I_0$ is the critical current of single junction. Thus, the effective inductance of the junction can be determined as

$$L_{jj}^* = \frac{\Phi_0}{2\pi I_0 \cos\left(\frac{\pi\Phi}{\Phi_0}\right)\cos\left(\delta_1 + \frac{\pi\Phi}{\Phi_0}\right)}. \quad (2)$$

This Josephson inductance $L_{jj}^*$ is tunable by a magnetic field. The character of magnetic field dependence for dc-SQUIDs coincides with the dependence for rf-SQUIDs. The measured transmission magnitudes through a coplanar waveguide containing single rf-SQUID and array of rf-SQUIDs depending on frequency and magnetic flux are presented in [13], [14]. The total flux $\Phi$ through the loop is related to the external flux $\Phi_{ext}$ and the current-induced flux by

$$\Phi_{ext} = \Phi + L_{geo} I , \quad (3)$$

where $I$ is the current through the loop. In addition to $L_{jj}^*$, the geometric inductance of the loop $L_{geo}$ contributes to the total inductance of the dc-SQUID. Using Kirchhoff laws we can present the dynamic equation for the flux threading the SQUID ring in the form

$$I_0 \cos\left(\frac{\pi\Phi}{\Phi_0}\right)\sin\left(\delta_1 + \frac{\pi\Phi}{\Phi_0}\right) + \frac{1}{R^*}\frac{d\Phi}{dt} + C^*\frac{d^2\Phi}{dt^2} + \frac{\Phi - \Phi_{ext}}{L_{geo}} = 0,$$
(4)

where Josephson phase difference is related to the flux $\Phi$ through the flux quantization condition $\delta_2 - \delta_1 = \frac{2\pi\Phi}{\Phi_0}$, $\Phi_0 = \frac{h}{2e} \cong 2.07 \times 10^{-15}$ Tm$^2$ is the magnetic-flux quantum, $h$ is Planck's constant, $e$ is the electronic charge. For a dc-SQUID without self inductance ($L_{geo}$=0) we can determine the maximal total Josephson current (supercurrent) $I_{jj\max} = 2I_0 \left|\cos\left(\frac{\pi\Phi_{ext}}{\Phi_0}\right)\right|$.

The dc-SQUID exhibits a resonant magnetic response at a resonant frequency

$$f_{mSQUID} = \frac{1}{2\pi\sqrt{\left(\frac{1}{L_{jj}^*} + \frac{1}{L_{geo}}\right)^{-1} C^*}} . \quad (5)$$

Let us note that for the choosing geometry of the MAs, the frequency of resonant magnetic response will depend on the self-inductance of the rods $L_r$ and can be written as

$$f_{mMA} = \frac{1}{2\pi\sqrt{\left(\frac{1}{L_{jj}^*} + \frac{1}{L_{geo}+L_r}\right)^{-1} C^*}} . \quad (6)$$

The planar dc SQUID array exhibits large magnetic response close to this resonance frequency. According to [8], we can write the relative magnetic permeability as

$$\mu_r = 1 + F\frac{\Phi_{res} - \Phi_{ext}}{\Phi_{ext}}, \quad (7)$$

where filling factor $F$ [10] depends on the geometry of the MA and spatial composition of the MMs. To neglect the magnetic interactions between individual SQUIDs, the parameter $F$ has to be very small ( $F \ll 1$ ). The analysis of eq.(7) demonstrates that depending on the value of $F$ and external flux, the relative magnetic permeability may take the negative value at frequencies $f > f_m$ [8].

It is necessary to note that for a case when 1D array of dc-SQUIDs is placed in the waveguide, it should be modeled as an equivalent lumped circuit coupled to the transmission line. The coupling is assumed to be pure magnetic with magnetic field perpendicular to the plane of the loop. For such model we should take into account a mutual inductance $M$ responsible for the coupling. Parameter $M$ should be found by considering the excitation of the SQUID by the magnetic field of electromagnetic wave [20]. For a transmission line of unit cell length $a$ the mutual inductance is equal to

$$M = \pi \frac{R_1^2 \mu_0}{a}, \quad (8)$$

where $\mu_0$ is the permeability of free space. Analysing the dispersion relation for the obtained system we can get the additional resonant frequency $f_t^2 = \frac{f_m^2}{1-q^2}$, where $q^2 = M^2/(LL_t)$, $L_t$ is the inductance per length $a$ of a transmission line, $L$ is a total inductance of the dc-SQUID. The frequency $f_t$ can be an upper bound of the band with negative magnetic permeability $\mu_r$.

### 2.2. Electric response of the system

In order to find the medium with simultaneously negative permittivity and permeability, we combine the dc-SQUIDs with superconducting rods. We assume that the rods with length $l_r$ and thickness $w_r$ ($w_r = R_1 - R_2$) can be characterized by a self-inductance $L_r$. For a transmission line periodically loaded by the rods we get a resonant electric response at the frequency

$$f_e = \frac{1}{2\pi\sqrt{L_r C_t}}. \quad (9)$$

Here $C_t$ is the capacitance per length $a$ of a transmission line.

### 3. Numerical simulations

The systematic numerical study of the dc-SQUID MMs design in the microwave part of the spectrum was performed to demonstrate the ability to produce LH behavior over a wide range of structure parameters. The numerical simulation was carried out by software package COMSOL Multiphysics. To get the resonant electric response in the range of magnetic MA tunability we optimize the critical current $I_0$. The simulations were done for the Nb superconducting rods and loops with Nb/AlO$_x$/Nb Josephson junctions. The area of JJ and the area of the capacitor yield nominal values of $I_0 = 0.97$ μA and $C_1 = C_2 = 0.32$ pF [15]. The composite medium is oriented in the waveguide. The incident rf power is fixed at -80dBm, $\Phi_{ext} = \Phi_0$. The incident wave propagating along the plane of the MAs` array excites a TEM mode in the waveguide with an rf magnetic field perpendicular to the plane of the SQUIDs. The transmission magnitude of the 4-elements' composite medium is illustrated in Fig.2. The transmission peak at frequency band between the resonant frequencies $f_{mMA}$ (magnetic resonance) and $f_e$ (electric resonance) corresponding to the NIM can be observed. We point out that $f_{mMA}$ and $f_e$ are sensitive functions of the length of the rods $l_r$. Increasing the length of the rods, one increases the rod self-inductance, thus, the resonant frequencies of the system go to lower values. Another observation that can be deduced from Fig.2 is that the distance between the resonant frequencies decreases with increasing of $l_r$. This fact is connected with the higher sensitivity of $f_e$ to this parameter. The resonance dips for all curves at frequencies about 12 GHz, 24 GHz and 36 GHz correspond to the cutoff frequencies of the waveguide.

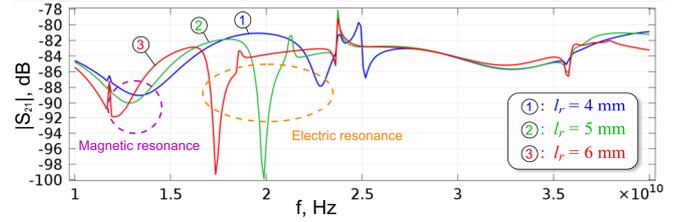

Figure 2: |S$_{21}$| of the 4 MAs array as a function of frequency at $w_r$=250 μm, $R_1$=400 μm, $R_2$=150 μm, $d$=1.2 mm

It is necessary to note that in our theoretical approach we did not take into account that the neighboring MAs are coupled to each other. The comparison of theoretical and numerical results indicates that the mutual inductance between the elements leads to the reduction of the resonant frequencies. Increasing the distance between the elements we decrease the mutual inductance. As the result for $d$=3.2 mm the frequency $f_{mMA}$ increases while the resonant frequency $f_e$ shifts to lower frequency range. The wider resonance dips are observed. For a case of uncoupled MAs, with increase of $d$ both resonant frequencies move to higher frequency range.

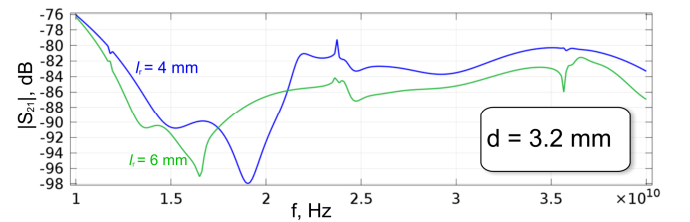

Figure 3: |S$_{21}$| of the 4 MAs array as a function of frequency at $w_r$=250 μm, $R_1$=400 μm, $R_2$=150 μm, $d$=3.2 mm

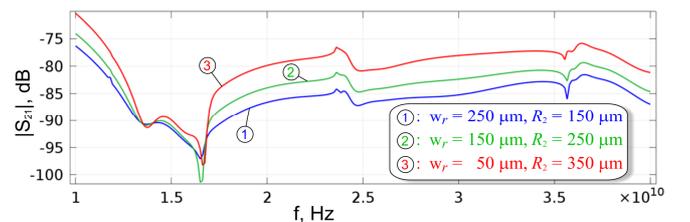

Figure 4: |S$_{21}$| of the 4 MAs array as a function of frequency at $l_r$=6 mm, $R_1$=400 μm, $d$=1.6 mm

Fig. 4 depicts the dependence of calculated transmission magnitude on the rod's width and internal radius of the SQUID. Let us note that the resonant frequencies are practically independent of these parameters. The highest transmission coefficient in NIM frequency band corresponds to the thinnest MAs.

To elucidate the effect of orientation of MAs on the transmission characteristics of the array we compare in Fig. 5 the transmission results of systems with identical and alternate orientation of MAs. As shown in Fig.5 the alternate orientation of MAs slightly affects the resonant frequency $f_e$ and leads to the higher transmission resonance for NIM.

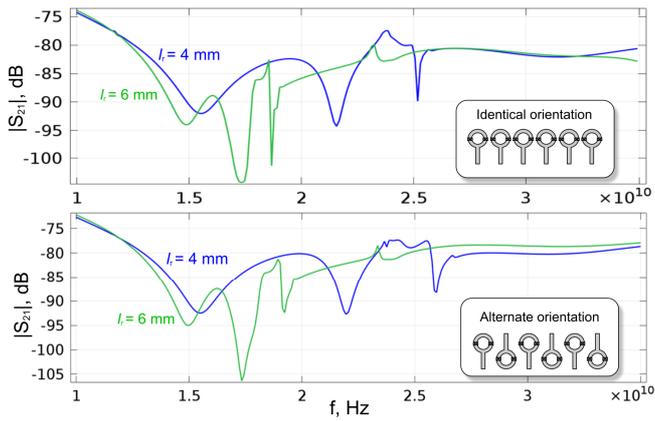

Figure 5: $|S_{21}|$ of the 6 MAs array as a function of frequency at $w_r$=250 μm, $R_1$=400 μm, $R_2$=150 μm, $d$=1.6 mm

## 4. Conclusions

We have analysed, both theoretically and numerically, a LH metamaterial composed of dc-SQUIDs with superconducting rods. The dependence of the electromagnetic wave propagation characteristics on the parameters of MAs is studied and discussed.

## Acknowledgements

The research work was partially supported by the European Union Seventh Framework Program (FP7-REGPOT-2012-2013-1) under grant agreement No. 316165. Partial support by the Ministry of Education and Science of Russian Federation in the framework of Increase Competitiveness Program of the NUST MISIS (contracts no. K2-2015-002, K2-2015-007, and K2-2016-051) is gratefully acknowledged.